\documentclass[aip,sd,graphicx,reprint]{revtex4-1}
\usepackage{amsfonts}
\usepackage{amsmath}
\usepackage{amssymb}
\usepackage{graphicx}
\usepackage{dcolumn}
\usepackage{bm}

\begin{document}

\title{Broadband plasmonic absorber for photonic integrated circuits}
\author{Xiao Xiong}
\author{Chang-Ling Zou}
\email{clzou321@ustc.edu.cn}
\author{Xi-Feng Ren}
\email{renxf@ustc.edu.cn}
\author{Guang-Can Guo}
\date{\today}

\affiliation{Key Lab of Quantum Information, University of Science and Technology of
China, Hefei 230026, P. R. China}

\begin{abstract}
The loss of surface plasmon polaritons has long been considered as a fatal
shortcoming in information transport. Here we propose a plasmonic absorber
utilizing this ``shortcoming" to absorb the stray light in photonic
integrated circuits (PICs). Based on adiabatic mode evolution, its
performance is insensitive to incident wavelength with bandwidth larger than
$300\:\mathrm{nm}$, and robust against surrounding environment and
temperature. Besides, the use of metal enables it to be very compact and
beneficial to thermal dissipation. With this $40\:\mathrm{\mu m}$-long
absorber, the absorption efficiency can be over $99.8\%$ at $1550\:\mathrm{nm%
}$, with both the reflectivity and transmittance of incident light reduced
to less than $0.1\%$. Such device may find various applications in PICs, to
eliminate the residual strong pump laser or stray light.
\end{abstract}

\pacs{}
\maketitle









Strong light is widely used in optical circuits: pumping the active medium
for light amplification and for coherent laser emission or the quantum dot
for single photon source, pumping nonlinear optical processes such as
four-wave mixing \cite{nonlinear}, and actuating and controlling the micro-
and nano-mechanical oscillators \cite{Tang,XGuo}. On the other hand, in
photonic integrated circuits (PICs), weak signal light or single photon are
used for classical and quantum information processing. Therefore, stray
light from strong pump/control light will induce errors in information
processing and high noise background. That is to say, an optical
\textquotedblleft dump\textquotedblright , which can perfectly absorb the
strong pump/control light with no reflection, is quite necessary in PICs.
During past decades, a variety of spatial two-dimensional (2D) or
three-dimensional (3D) absorbers composed of array of plasmonic
nano-resonators have been proposed \cite{Abs1,Abs6,Abs2,Abs3,Abs4,Abs5}.
Surprisingly, however, integrated absorber compatible with dielectric
waveguide has been omitted.

One may borrow the idea from the spatial absorbers to PICs by using
critically coupled resonators to absorb the incoming light with neither
reflection nor transmission. However, there exists several limitations in
resonator-based absorber. Firstly, the cavity has to work under the critical
coupling condition, which is highly dependent on the geometric parameters of
cavity. Secondly, the absorption spectrum is a Lorentz-shaped profile with a
relatively narrow bandwidth. Thirdly, the resonance wavelength is sensitive
to temperature, thus the absorption spectrum will shift due to light
heating, and lead to instability. Fourthly, finite cavity resonance also
corresponds to limited response speed, which means laser pulse can not be
absorbed effectively before stable field is established in the cavity. In
addition, the resonances are polarized in PICs, hence the two orthogonal
polarizations usually cannot be absorbed simultaneously.

In this Letter, we propose a broadband plasmonic absorber for PICs. Since
surface plasmon polaritons (SPPs) enable extreme confinement and have big
inherent losses in propagation \cite{SPP1,SPP2}, they are of course the best
candidate to fulfill the task for absorption in PICs. However, the impedance
mismatch between dielectric and plasmonic modes will inevitably lead to
reflection. Therefore, we utilize the adiabatic mode evolution to
efficiently convert the dielectric waveguide mode to plasmonic mode \cite%
{adiabatic}. As a result, the absorption can be achieved over a wide range
of wavelength, and remains robust against surrounding environment and
temperature \cite{polarizer}. By a simple circular arc structure of dozens
of microns long, we realized a waveguide-integrated adiabatic absorber,
whose absorption efficiency can reach more than $99.8\%$ at $1550\:\mathrm{nm%
}$, with both the reflectance and transmittance reduced to less than $0.1\%$%
. And the absorption efficiency keeps over $99.5\%$ from $1400\:\mathrm{nm}$
to $1700\:\mathrm{nm}$ for both horizontal ($H$) and vertical ($V$)
polarizations. Additional to the excellent optical performance, the absorber
has very good thermal conductivity considering that laser heating can be
conducted by metal wires or thermal antenna. We believe this compact,
broadband, robust and thermal-conductive absorber may find applications in
practical PICs.

\begin{figure}[tbp]
\centerline{ \includegraphics[width=8.5cm]{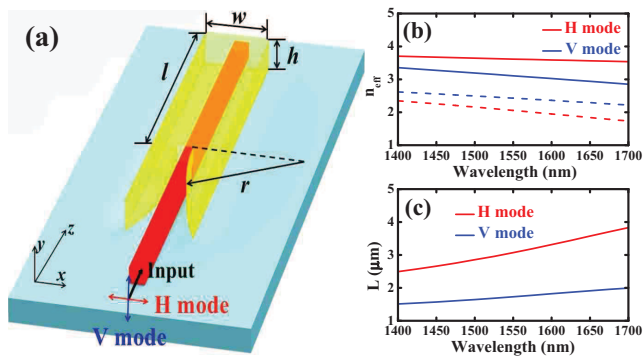}}
\caption{(a) Schematic of the adiabatic absorber, which consists of the
adiabatic circular arc (radius $r$) and the absorption rectangle (size $%
l\times w\times h$). Incident light includes two orthogonal polarizations ($%
H $ and $V$ modes) and propagates along the $z$-axis. (b) Dependence of
effective refractive index $n_{eff}$ on incident wavelength $\protect\lambda$
for dielectric $H/V$ modes (dashed red/blue curves) and plasmon $H/V$ modes
(solid red/blue curves), respectively. Here, $l=20\:\mathrm{\protect\mu m}$,
$w=1.5\:\mathrm{\protect\mu m}$, $h=0.8\:\mathrm{\protect\mu m}$. (c)
Dependence of propagation length $L$ on $\protect\lambda$ for plasmon $H$
(red curve) and $V$ (blue curve) modes, with $w=1.5\:\mathrm{\protect\mu m}$
and $h=0.8\:\mathrm{\protect\mu m}$.}
\end{figure}

The schematic illustration of our proposal is shown in Fig. 1(a). A silicon
waveguide is fabricated on silica substrate, with the waveguide
cross-section being $300\:\mathrm{nm}\times400\:\mathrm{nm}$. And gold (Au)
of height $h$ covers the end of the waveguide and consists of two parts: the
tangent circular arc with radius of curvature being $r$, and the rectangle
with width $w$ and length $l$. The working wavelength is focused on the
telecommunication band around $\lambda=1550\ \mathrm{nm}$. The refractive
indices of silicon and silica are $n_{Si}=3.5$ and $n_{SiO_{2}}=1.444$,
respectively. And the permittivity of Au is calculated with Lorentz-Drude
model according to parameters from Ref. \cite{constant}. Then we performed
numerical simulations with COMSOL Multiphysics 4.3 to test the performance
of this absorber. It's obvious that the performance of an absorber should be
estimated with two parameters: transmittance $T$ and reflectivity $R$. Here,
$T$ is defined as the ratio of total outgoing energy flux to total incident
energy flux \cite{reflect}, and $R$ is the ratio of backward-propagating
intensity to forward-propagating intensity.

Firstly, we calculated the effective refractive index $n_{eff}$ of
dielectric eigenmode (in pure silicon waveguide without metal) and plasmon
eigenmode (in metal-covered silicon waveguide) as a function of wavelength $%
\lambda $, as shown in Fig. 1(b). We observed that the difference between
dielectric and plasmon modes is rather big for both polarizations,
especially for $H$ polarization. This big difference, which indicates the
mode mismatching between dielectric and plasmon modes, will induce large
reflectivity on the interface between pure silicon waveguide and
metal-covered silicon waveguide. In order to reduce the reflectivity, a
designed taper can be used to convert dielectric guided modes to plasmon
modes adiabatically \cite{Taper,taper}. Here, we just adopted a circular arc
which is quite simple but works well.

Then, we calculated the propagation length of plasmon modes $L$ depending on
$\lambda$ for $H$ and $V$ modes, respectively, as displayed in Fig. 1(c). As
the incident wavelength increases, the propagation length $L$ increases for
both polarizations. And the propagation length $L$ of $H$ mode keeps larger
than that of $V$ mode. Nevertheless, for a wide bandwidth of $300\ \mathrm{nm%
}$, the propagation length $L$ of both $H$ and $V$ modes are below $4\
\mathrm{\mu m}$. According to $I(l)=I_{0}e^{-l/L}$, we get to know that
longer $l$ will bring better absorption performance. Especially, at $\lambda
=1550\ \mathrm{nm}$, the propagation length of $H$ mode is larger, being $%
L=3\ \mathrm{\mu m}$. Thus, we fixed $l$ to be $l=20\ \mathrm{\mu m}$,
expecting an absorption efficiency of $I(l)/I_{0}=e^{-20/3}\approx 0.1\%$.
In addition, we also calculated the dependence of propagation length $L$ on $%
w$ and $h$ for $H$ and $V$ modes. It increases with increasing $w$ and $h$
firstly, and get saturated over certain value ($w>0.5\ \mathrm{\mu m}$, $%
h>0.5\ \mathrm{\mu m}$). This is because energy is mostly confined in the
metal when the Au coating is ultra-thin, rather than in the waveguide. When $%
w>0.5\ \mathrm{\mu m}$ and $h>0.5\ \mathrm{\mu m}$, the energy of
propagating mode is mostly confined in the dielectric waveguide, then the
absorption of this device won't be sensitive to the size of metal coating.
So we fixed $w=1.5\ \mathrm{\mu m}$ and $h=0.8\ \mathrm{\mu m}$ in the
saturated region in our studies.

It is conceivable that if the conversion from dielectric mode to plasmon
mode in the first half of this absorber is more adiabatic, the reflectivity
due to momentum mismatching will be smaller \cite{adiabatic}. Similarly, if
the plasmon mode propagates farther, the loss of incident light due to
metallic absorption is larger. That is to say, larger $r$ brings smaller $R$
while longer $l$ brings smaller $T$. Next, we'll confirm our predictions
with 3D numerical simulations. Matched fundamental modes with two orthogonal
polarizations ($H$ and $V$ modes) are incident at the waveguide facet,
respectively, and propagate along $z$ direction, with the coordinate origin
located at the waveguide center of incident plane. Note that, benefiting
from the propagation losses in metal, $T$ can always be reduced as small as
possible, as long as $l$ is large enough. As a result, we focus on the
influence of adiabatic radius $r$ on reflectivity $R$ in the following.

\begin{figure}[tbp]
\centerline{\includegraphics[width=8cm]{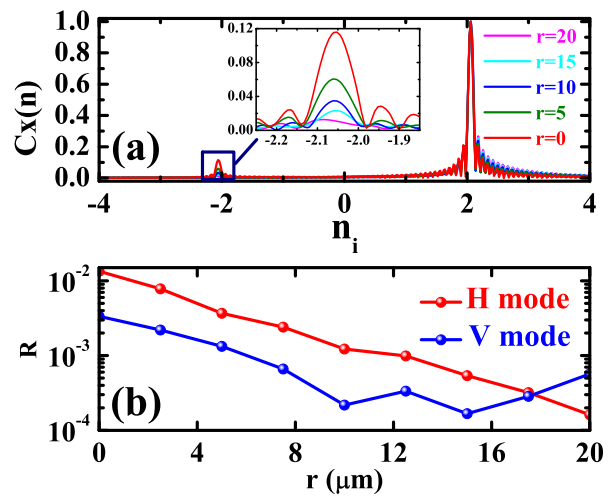}}
\caption{(a) Fourier-transformed spectrum of electric field component $%
E_{m}(z)$ in the adiabatic part, with $r=0$, $5$, $10$, $15$, $20$,
respectively. Here, the incident field is $H$ mode, with $l=20$, $w=1.5$, $%
h=0.8$ (in the unit of $\mathrm{\protect\mu m}$). Inset: enlarged view of
the peak which stands for the reflected field. (b) Dependence of
reflectivity $R$ on $r$ for $H$ and $V$ modes, respectively.}
\end{figure}

Denote the electric field component along $m$ direction ($m=x(y)$ for $H(V)$
polarization) as $E_{m}(z)$, which is distributed along the center line of
waveguide as a function of $z$. In the adiabatic part, $E_{m}(z)$ should be
a sum of forward- and backward-propagating dielectric modes as

\begin{equation}
E_{m}(z)=\sum_{n_{i}}C_{m}(n_{i})e^{-jn_{i}kz},  \label{Ez}
\end{equation}%
where $n_{i}$ is the effective refractive index of different eigenmodes, $%
C_{m}(n_{i})$ is the corresponding mode amplitude, and $k=\frac{2\pi }{%
\lambda }$ is the wave number. By applying Discrete Fourier Transform (DFT)
to $E_{m}(z)$, the coefficient $C_{m}(n_{i})$ can be expressed as
\begin{equation}
C_{m}(n_{i})=\frac{k}{2\pi }\sum_{z}E_{m}(z)e^{jn_{i}kz}.  \label{Cn}
\end{equation}%
And the sign before $n_{i}$ represents exactly the energy propagation
direction. As shown in Fig. 2(a), the Fourier-transformed spectrum consists
of two main peaks that are symmetric distributed. The right main peak stands
for the forward-propagating incident mode with $n=+n_{i}$, while the left
one is the backward-propagating reflected mode with $n=-n_{i}$. Here, the
priminent peak for $n<0$ corresponds to reflected fundamental mode. Then,
the reflectivity $R$ is obtained as
\begin{equation}
R=\left\vert \frac{C_{m}(-n_{i})}{C_{m}(+n_{i})}\right\vert ^{2}.  \label{R}
\end{equation}%
Here, the incident field has been normalized to be unit for simplicity.
Besides, it's worth noting that there are many side lobes around the main
peaks, which originates from the finite sampling interval. Furthermore, the
dependences of $R$ on the arc radius $r$ for different polarization are
plotted in Fig. 2(b). As we can see, for $r>15:\mathrm{\mu m}$, the
reflectivity $R$ are reduced to be less than $0.1\%$ for both $H$ and $V$
modes.

\begin{figure}[tbp]
\centerline{\includegraphics[width=7cm]{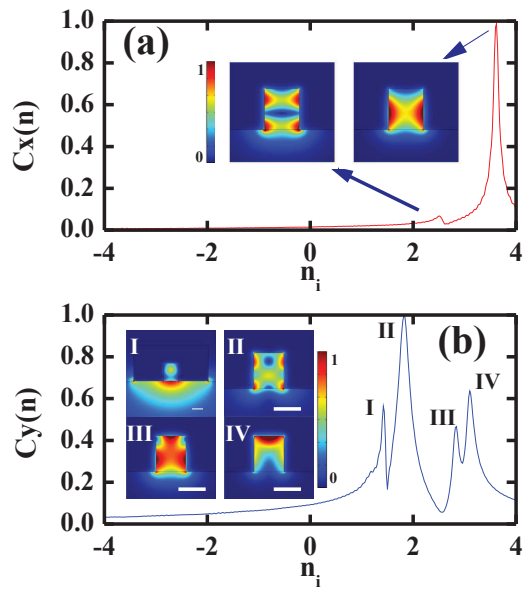}}
\caption{Normalized Fourier-transformed spectra of electric field component $%
E_{m}(z)$ in the absorption part for (a) $H$ mode and (b) $V$ mode,
respectively, with $r=20\:\mathrm{\protect\mu m}$. Inset: the field
distributions of plasmon eigenmodes, which correspond to the peaks in
Fourier-transformed spectra, respectively. Scale bar in the insets I-IV: $%
300\:\mathrm{nm}$.}
\end{figure}

Similarly, by applying DFT to $E_{m}(z)$ in the absorption part, we obtained
the Fourier-transformed spectra of $H$ and $V$ modes, which contain the
information about plasmon eigenmodes. As illustrated in Fig. 3(a), the
normalized spectrum for $H$-polarized incident field consists of two peaks.
While no peak appears in the region $n_{i}<0$, it means that there are no
reflected modes in the absorption part of this absorber. As for the
effective refractive indices corresponding to the two peaks, they are
exactly consistent with the $H$-polarized plasmon eigenmodes calculated
under 2D simulations (shown in the inset in Fig. 3(a)), with the major
(minor) peaks being plasmon fundamental (high-order) eigenmode. Unlike the
situation in Fig. 3(a), the normalized spectrum for $V$-polarized incident
field (Fig. 3(b)) includes four peaks, and they stand for the four $V$%
-polarized plasmon eigenmodes of different order (inset of Fig. 3(b)),
respectively. Here, many $V$-polarized plasmon eigenmodes of higher-order
are excited, due to the discontinuity of metal in the interface between the
arc and rectangle at the top of waveguide for $V$ polarization. This sudden
change of refractive index also results in the special trend of the blue
curve shown in Fig. 2(b), where $R$ of $V$ mode does not decrease
monotonously with the increase of $r$. While for $H$ polarization, the
fundamental dielectric mode is adiabatically converted into fundamental
plasmon mode (with the high-order plasmon mode being negligible), and the
reflectivity $R$ drops monotonously as $r$ increases, since the metal is
getting close to the waveguide slowly enough. Therefore, the performance of $%
V$ mode can be improved by etching or evaporating a metal slope in the
interface between the arc and rectangle at the top of waveguide.

\begin{figure}[tbp]
\centerline{\includegraphics[width=7.5cm]{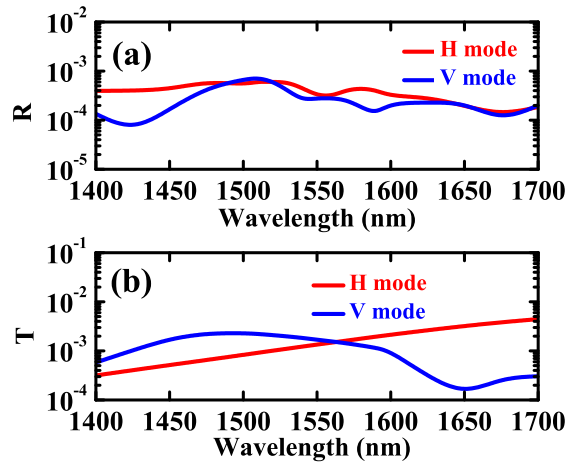}}
\caption{Dependence of (a) $R$ and (b) $T$ on $\protect\lambda$ for $H$ and $%
V$ modes. Here, $l=20$, $w=1.5$, $h=0.8$, $r=17.5$ (in the unit of $\mathrm{%
\protect\mu m}$).}
\end{figure}

Finally, we calculated the dependence of $R$ and $T$ on incident wavelength $%
\lambda$ for both $H$ and $V$ polarizations. As shown in Fig. 4(a), the
reflectivity $R$ is kept less than $0.1\%$ for both polarizations over a
bandwidth as wide as $300\:\mathrm{nm}$. As for the transmission $T$ shown
in Fig. 4(b), it increases with increasing incident wavelength for $H$
polarization, while maintains well below $0.3\%$ over a bandwidth of $300\:%
\mathrm{nm}$ for $V$ polarization. In order to further increase the
absorption efficiency, the absorption part can be engineered to be longer,
which is quite convenient. Or, the waveguide in the absorption part can be
tapered to adiabatically transfer more energy into the metal and improve the
absorption loss.

There are still several points that we want to discuss. For instance, to
further improve the absorber's performance with less reflectivity as well as
less transmission, the process of mode conversion can be engineered to be
more adiabatic (larger $r$), and the distance for absorption needs to be
longer (larger $l$). And considering the adoption of metal, which is a good
thermal conductor, heat dissipation can be easily improved by adding more
metal, etching nanostructures at the top of the metal rectangle like heat
sink, and guiding the thermal energy with a metallic wire or thermal emitter
\cite{emitter}. Since incident optical energy is totally converted to
thermal energy, it fundamentally eliminates the negative effect of stray
light. While for resonance-based absorber, a portion of the incident light
will be transformed to scattering light. Additionally, since the imaginary
part of the refractive index of Au increases with higher temperature, the
increase of temperature in metal caused by energy absorption will further
improve the absorption efficiency of this absorber, rather than wrecking it.
Finally, since this absorber works based on adiabatic mode conversion, its
performance is insensitive to geometric parameters and imposes little
requirement on processing technology. Our calculation proves that, even if
the lateral displacement between dielectric waveguide and plasmonic absorber
is as large as $100\:\mathrm{nm}$, the absorption efficiency for both $H$
and $V$ modes can still be maintained around $99\%$.

In conclusion, we propose a plasmonic absorber for guided waves in PICs,
which is based on adiabatic mode conversion and can achieve an absorption
efficiency more than $99.8\%$ (with both reflectivity and transmission less
than $0.1\%$). And the use of metal enables the device to be very compact ($%
1\:\mathrm{\mu m}\times1\:\mathrm{\mu m}\times40\:\mathrm{\mu m}$) and
conducive to heat dissipation. Besides, since the absorption is realized
adiabatically with a circular arc, this absorber has a bandwidth as wide as $%
300\:\mathrm{nm}$ for both $H$ and $V$ polarizations, and its performance is
also robust against surrounding environment and temperature. It is essential
for integrated photonic chips, and will find applications in PICs-based
optics.

\begin{acknowledgments}
This work was funded by NBRP (grant nos. 2011CBA00200 and 2011CB921200),
the Innovation Funds from the Chinese Academy of Sciences (grant no.
60921091), NNSF (grant nos. 10904137, 10934006 and 11374289), the
Fundamental Research Funds for the Central Universities (grant no.
WK2470000005), and NCET.
\end{acknowledgments}

%
%

%


\bibliographystyle{plain}
\bibliography{your-bib-file}

\end{document}